\pdfoutput=1

\documentclass[11pt]{article}

\usepackage[]{ACL2023}

\usepackage{times}
\usepackage{latexsym}

\usepackage[T1]{fontenc}

\usepackage[utf8]{inputenc}

\usepackage{microtype}

\usepackage{inconsolata}

\usepackage{graphicx}
\usepackage{subfigure}
\usepackage{amsmath}
\usepackage{amssymb}

\usepackage{algorithm}
\usepackage{algorithmic}

\usepackage{booktabs}
\usepackage{multirow}
\usepackage{makecell}

\usepackage{CJKutf8}
\title{Semantic Steganography: A Framework for Robust and High-Capacity Information Hiding using Large Language Models}

\author{Minhao Bai$^\diamondsuit$, Jinshuai Yang$^\diamondsuit$, Kaiyi Pang$^\diamondsuit$, Yongfeng Huang$^{\diamondsuit,\spadesuit}$, Yue Gao$^\spadesuit$\\
    $^\diamondsuit$Tsinghua University, $^\spadesuit$Zhongguancun Laboratory
}

\begin{document}
\maketitle
\begin{abstract}
In the era of Large Language Models (LLMs), generative linguistic steganography has become a prevalent technique for hiding information within model-generated texts. However, traditional steganography methods struggle to effectively align steganographic texts with original model-generated texts due to the lower entropy of the predicted probability distribution of LLMs. This results in a decrease in embedding capacity and poses challenges for decoding stegos in real-world communication channels. \\
To address these challenges, we propose a semantic steganography framework based on LLMs, which constructs a semantic space and maps secret messages onto this space using ontology-entity trees. This framework offers robustness and reliability for transmission in complex channels, as well as resistance to text rendering and word blocking. Additionally, the stegos generated by our framework are indistinguishable from the covers and achieve a higher embedding capacity compared to state-of-the-art steganography methods, while producing higher quality stegos.
\end{abstract}

\section{Introduction}

In the age of information, private communication and freedom of speech are threatened by widespread surveillance and social media censorship. This infringement on fundamental human rights has eroded the privacy protections in liberal democracies. Social media platforms, once seen as promoters of free expression, are now accused of silencing dissenting voices. To counter this, individuals are turning to covert communication technologies to protect their privacy and freedom of speech. Steganography, as a widely used way of constructing covert communication to avoid arousing suspicion, involves embedding private messages within seemingly ordinary carriers, making it difficult for censors to detect and block them.

With the rapid iterations of Large Language Models (LLMs) \cite{LLAMA,ChatGLM}, texts generated by LLMs flood cyberspace, providing a thriving environment for generative linguistic steganography \cite{HC1,HC2,HC3,AC,SAAC,METEOR,MEC,DISCOP,RNNstega,VAEstega,NewBinstega,ICstega}. As a technique of modifying the generation process of language models to embed secret messages, mainstream steganography methods \cite{METEOR,MEC,DISCOP} focus on aligning steganographic texts (stegos for short) with original model-generated texts (covers for short). 
However, current steganography techniques have two major weak points.

\begin{figure}[ht]
    \subfigure[AC + ChatGLM2-6B-int4]{
		\includegraphics[width=0.5\textwidth]{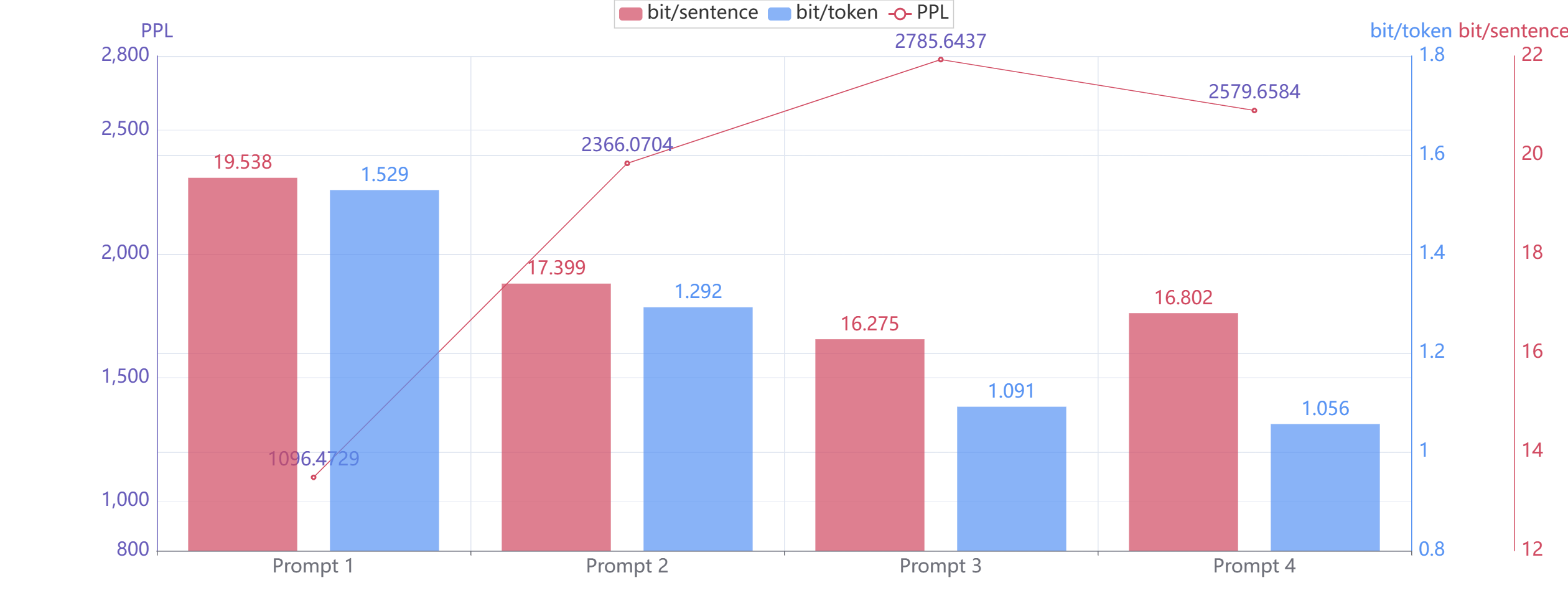}
		\label{AC + ChatGLM2-6B-int4}
    }
    \subfigure[AC + ChatGLM2-6B]{
		\includegraphics[width=0.5\textwidth]{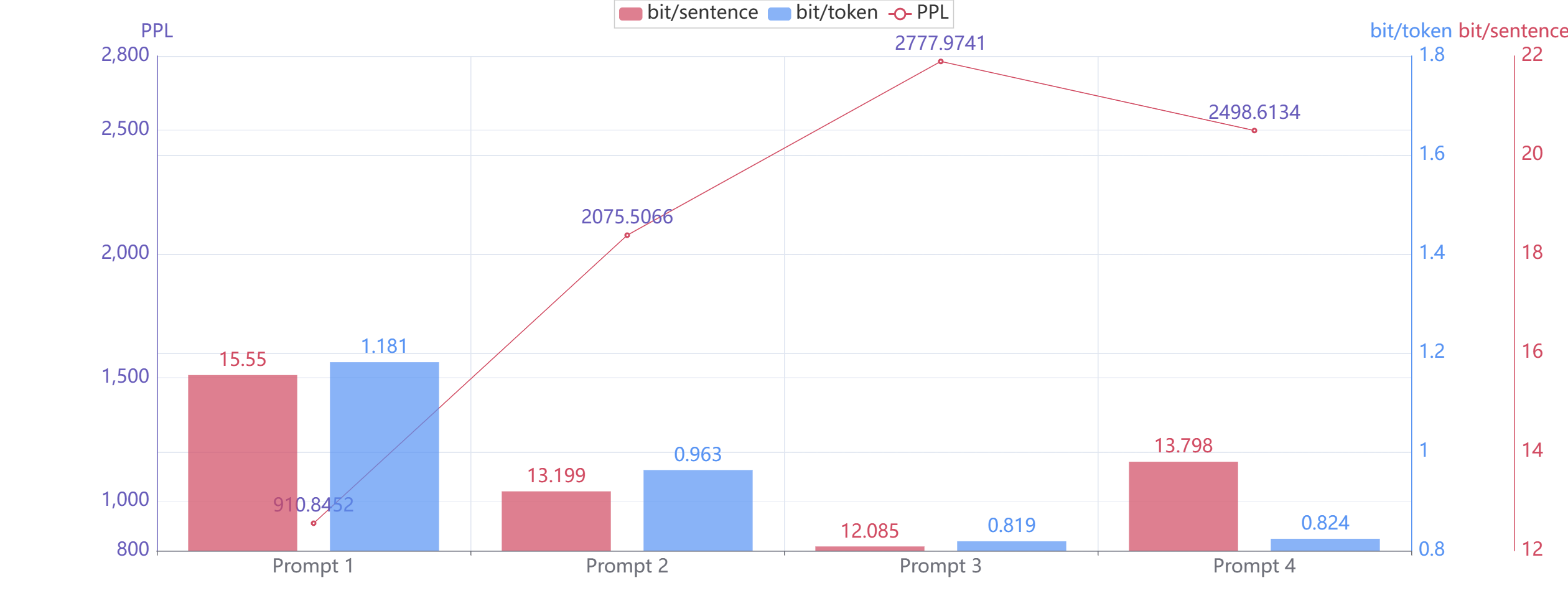}
		\label{AC + ChatGLM2-6B-int4}
    }
    \caption{Perplexity(PPL) and embedding rate of stegos generated by Arithmetic Coding(AC)\cite{AC} with ChatGLM2-6B-int4/ChatGLM2-6B. The left axis represents PPL, while the right two axes represent embedding rate, estimated in bits per sentence and bits per token. Prompt 1 is null, while prompts 2-4 require the model to generate a single given word, namely ``the'', ``like'', or ``Washington'', respectively.}
    \label{fig:baseline ppl and embedding rate}
\end{figure}

{\bf Low Symbol-level Entropy.} Given the same text prefix, the entropy of the predicted probability distribution of LLMs is likely to be lower than that of GPT-2 \cite{GPT2} or BERT \cite{BERT}. For SOTA provably secure steganography algorithms \cite{METEOR,MEC,DISCOP}, entropy is an upper bound on the embedding capacity. A large decrease in entropy leads to a dramatic decrease in embedding capacity. Nevertheless, it seems that the more powerful models have lower entropy. As figure \ref{fig:baseline ppl and embedding rate} shows, with the same steganography method Arithmetic Coding (AC) \cite{AC}, the embedding rate of ChatGLM-2-6B is about 1/4 $\sim$ 1/5 lower than that of ChatGLM-2-6B-int4. With stricter top-k truncation and more detailed prompting, the embedding rate may decrease further.

{\bf Not Robust.}
When applying these steganography methods to real-world communication channels, particularly in social networks, we have found that most received stegos cannot be decoded. This problem may be caused by text transcoding, word blocking, or ambiguity of tokenizer, resulting in token insertion, deletion, or replacement. 




The challenges of applying state-of-the-art steganography methods to LLMs highlight the lack of robustness and capacity of symbolic token-level embedding. 
Therefore, we propose a semantic steganography framework based on LLMs. This framework primarily constructs a semantic space and uses LLMs to generate responses that belong to a subset of that space. To ensure the quality of LLMs' output and their embedding capacity, we mapped the secret messages onto the semantic space using ontology-entity trees. During the decoding phase, the semantic information in stegos will be retrieved and converted back into secret messages.

Compared to the symbol-based steganography techniques, our framework has the following advantages:
\begin{itemize}
    \item Our framework is more reliable and robust for transmission in network environments. The stegos generated by our method are able to resist ambiguous tokenizing and text transcoding.
    \item The stegos in our framework are directly generated by LLMs, which is different from current steganography works that manually modify the generation procedure. From the semantic aspect, our method produces almost indistinguishable semantic contents.
    \item Our framework achieves a higher embedding capacity than state-of-the-art steganography methods under the same settings while producing higher-quality stegos.
\end{itemize}

\section{Methods}
\subsection{Construct the Semantic Space}

Semantic space is a set in which sentences are represented based on their meanings and relationships. And the basic step of our steganography framework is to construct the semantic space.

Various methods exist for constructing a semantic space.

(1) {\bf Classifiers.} In previous work \cite{semantic1}, classifiers were used to control the semantic information at the sentence level, but such classifiers need training and are not easy to share with the receiver. To ensure objectivity, we avoid using emotions or main themes as they are not realistic due to limited semantic space and restricted embedding capacity. Additionally, the meanings inside the sentence are mostly unused.  

(2) {\bf Embeddings.} The embedding output of language models can be used to construct a semantic space, but this method seems to be too sensitive and difficult to design. While this does not affect the encoding method, it can confuse the decoding process. We believe that using the embedding output of language models is feasible and requires further exploration.

(3) {\bf Entities.} Entities are considered to be effective and efficient for steganography encoding. The capacity of steganography is associated with the number of entities, because the more entities we have, the more bits we can use to uniquely represent each entity. So adding more entities is a feasible and convenient approach to expanding capacity. 




We prefer to use \textbf{Entities} to construct the semantic space because there exists a helpful structure called the ontology-entity tree. This tree comprises multiple top layers of concepts and the final layer of entities. The paths within this tree indicate a process of gradual refinement from a vague concept to a specific entity, offering additional information to describe the final entity.

Open-sourced ontology-entity trees do not typically contain information about the frequency of these entities or the relationships between them. This information is essential for estimating semantic distribution. Therefore, we embarked on constructing our own ontology-entity tree.

We extracted entities using the PaddleNLP UIE model \cite{paddlenlp}. The dataset used for extraction and construction of the semantic space is LCCC \cite{LCCC}, a large-scale cleaned dataset containing 12 million daily conversations. Based on the extraction results, we hand-crafted two upper layers of concepts and a bottom layer of entities to construct the ontology-entity tree. The first level of the tree includes fundamental concepts such as ``person'' and ``location''. The subsequent level comprises subconcepts like ``tourism location'' or ``educational location''. The final level contains entities such as ``Las Vegas'' or ``Taj Mahal'', which belong to the subconcept ``tourism location''.

This tree also provides additional information to assist language models in generating decodable responses and determining which entity to use. The model may get confused when the entity ``Washington" is given, since it could represent a person or a location. But if we use the path from the root of the tree to the leaf node of the entity, we can get a detailed entity like ``Location/Tourism Location/Washington". Therefore the words that have multiple meanings can be distinguished and correctly extracted. 



For any entity $e_i \in \mathcal{E}$, we construct an extraction method $Ext_{e_i}$. Using the extraction method $Ext_{e_i}$ we can extract the number of entity $e_i$ that appears in a sentence, denoted as $Ext_{e_i}(S) = n_i$. This extraction method can be completed by LLMs with appropriate prompts or other machine learning modules.

We define the {\bf type} of a sentence as follows:
The {\bf type} of sentence $S$ is $e_1^{n_1} ... e_{|\mathcal{E}|}^{n_{|\mathcal{E}|}}$, where $e_i \in \mathcal{E}$ is an entity, $n_i = Ext_{e_i}(S)$ is the times that $e_i$ appears in sentence $S$.
For instance, consider the sentence ``An apple a day, keeps the doctor away'' with the entities ``apple'' and ``doctor''. From this, we can determine that the {\bf type} of this sentence is $\text{apple}^1\text{doctor}^1$.
We define the length of a {\bf type} $|T|$ as the number of entities inside the sentence.
\begin{equation}
    |T| = \sum_{i = 1}^{|\mathcal{E}|} n_i
\end{equation}

For the sake of clarity, we provide a definition of the partial order relation between types: \textbf{type} $T^{(1)} = e_1^{n^{(1)}_1} \cdot\cdot\cdot e_{|\mathcal{E}|}^{n^{(1)}_{|\mathcal{E}|}}$ is not greater than \textbf{type} $T^{(2)} = e_1^{n^{(2)}_1}\cdot\cdot\cdot e_{|\mathcal{E}|}^{n^{(2)}_{|\mathcal{E}|}}$ if and only if

\begin{equation}
    \forall i \in \{1,2,\cdot\cdot\cdot,|\mathcal{E}|\}, n^{(1)}_i \leq n^{(2)}_i
\end{equation}

We can also define an add operation on the \textbf{type}, which represents combining 2 sentences into one.
\begin{equation}
    T^{(1)} + T^{(2)} = e_1^{n^{(1)}_1 + n^{(2)}_1} \cdot\cdot\cdot e_{|\mathcal{E}|}^{n^{(1)}_{|\mathcal{E}|}+n^{(2)}_{|\mathcal{E}|}}
\end{equation}


Sentences with the same type are highly correlated as they are likely referring to the same entities and have a relationship. We define {\bf class} $\mathcal{C}$ to denote the set of possible sentences that share the same {\bf type}. 

\begin{equation}
    \mathcal{C}(e_1^{n_1} \cdot\cdot\cdot e_{|\mathcal{E}|}^{n_{|\mathcal{E}|}}) = \{S|\textbf{type}(S) = e_1^{n_1} \cdot\cdot\cdot e_M^{n_M}\}
\end{equation}

In the end, the semantic space is defined as the set of all possible classes.

\begin{equation}
    \mathcal{S} = \{\mathcal{C}(e_1^{n_1} \cdot\cdot\cdot e_{|\mathcal{E}|}^{n_{|\mathcal{E}|}})| e_i \in \mathcal{E}, n_i \in \mathbb{N_+}\}
\end{equation}



Instead of generating a sentence with specific attributes, we prefer to determine and arrange the entities that should appear in the output.

\begin{algorithm}[t]

\renewcommand{\algorithmicrequire}{\textbf{Input:}}
\renewcommand{\algorithmicensure}{\textbf{Output:}}
\caption{Estimate the probability of each node}
\label{alg:assign}
\begin{algorithmic}[1]
\REQUIRE Ontology-entity tree $\mathcal{T}_o$, type prefix $T_{pre}$, empirical distribution $p(\mathcal{C}({T}))$, entity set $\mathcal{E}$
\ENSURE Probabilities of nodes $P(\cdot)$  
\STATE $\text{$\cdot\cdot\cdot$ Initial probabilities of nodes in tree}$
\FOR {$node \in \mathcal{T}_o$}
        \STATE $P(node)\leftarrow0$
\ENDFOR
\STATE $\text{$\cdot\cdot\cdot$ Assign probabilities to entities}$
\STATE $sum \leftarrow 0 $
\FOR {$e \in \mathcal{E}$}
    \FOR { $\mathcal{C}({T}) \in \mathcal{S} \quad\AND\quad p(\mathcal{C}) \neq 0$}
        \IF {$T_{pre}+e \leq T$} 
            \STATE $P(e) \leftarrow P(e) + p(\mathcal{C}({T})) $
        \ENDIF
    \ENDFOR
    \STATE $sum \leftarrow sum + P(e)$
\ENDFOR
\STATE $\text{$\cdot\cdot\cdot$ Assign probabilities to stop sampling}$
\STATE $P(stop) \leftarrow p(\mathcal{C}({T_{pre}}))$
\STATE $sum \leftarrow sum + p(\mathcal{C}({T_{pre}}))$
\STATE $\text{$\cdot\cdot\cdot$ Normalization}$
\FOR {$e \in \mathcal{T}_o$}
    \STATE $P(e) \leftarrow P(e) /sum$
\ENDFOR
\STATE $\text{$\cdot\cdot\cdot$ Accumulate probabilities to upper nodes}$
\FOR {$e \in \mathcal{T}_o$}
    \STATE $parent\leftarrow e.parent$
    \WHILE{$parent \neq T.ROOT$}
        \STATE $P(parent) \leftarrow P(parent) + P(e)$
    \ENDWHILE
\ENDFOR
\RETURN  $P(\cdot)$
\end{algorithmic}
\end{algorithm}

\begin{algorithm}[ht]
\renewcommand{\algorithmicrequire}{\textbf{Input:}}
\renewcommand{\algorithmicensure}{\textbf{Output:}}
\caption{Sample a type from semantic space}
\label{alg:sample}
\begin{algorithmic}[1]
\REQUIRE Ontology-entity tree $\mathcal{T}_o$, empirical distribution $p(\mathcal{C}({T}))$, cipher bits $C$, PRF $F_{key}$
\ENSURE Target type  $T_t$
\STATE $\text{$\cdot\cdot\cdot$ Randomize cipher bits}$
\STATE $B \leftarrow F_{key}(C)$

\STATE $T_t \leftarrow null$
\STATE $\text{$\cdot\cdot\cdot$ Select entities one by one}$
\WHILE {$pointer \neq stop$}
\STATE $sum \leftarrow 0$
\STATE $\text{$\cdot\cdot\cdot$ Select nodes layer by layer}$
\STATE $pointer \leftarrow ROOT$
    
    \FOR {$node \in pointer.child$}
        \IF {$sum + P(node) \geq  \sum_{i = 1}^{n} B[i]*2^{-i}$}
            \STATE $pointer \leftarrow node$\\
            \STATE $B \leftarrow B[n:]$
            \STATE \textbf{break}
        \ENDIF
        \STATE $sum \leftarrow sum + P(node)$
    \ENDFOR
\STATE $T_t \leftarrow T_t + pointer$
\ENDWHILE
\end{algorithmic}
\end{algorithm}

\subsection{Sample from the Semantic Distribution}
\begin{figure*}[ht]
\centering
\includegraphics[scale=0.45]{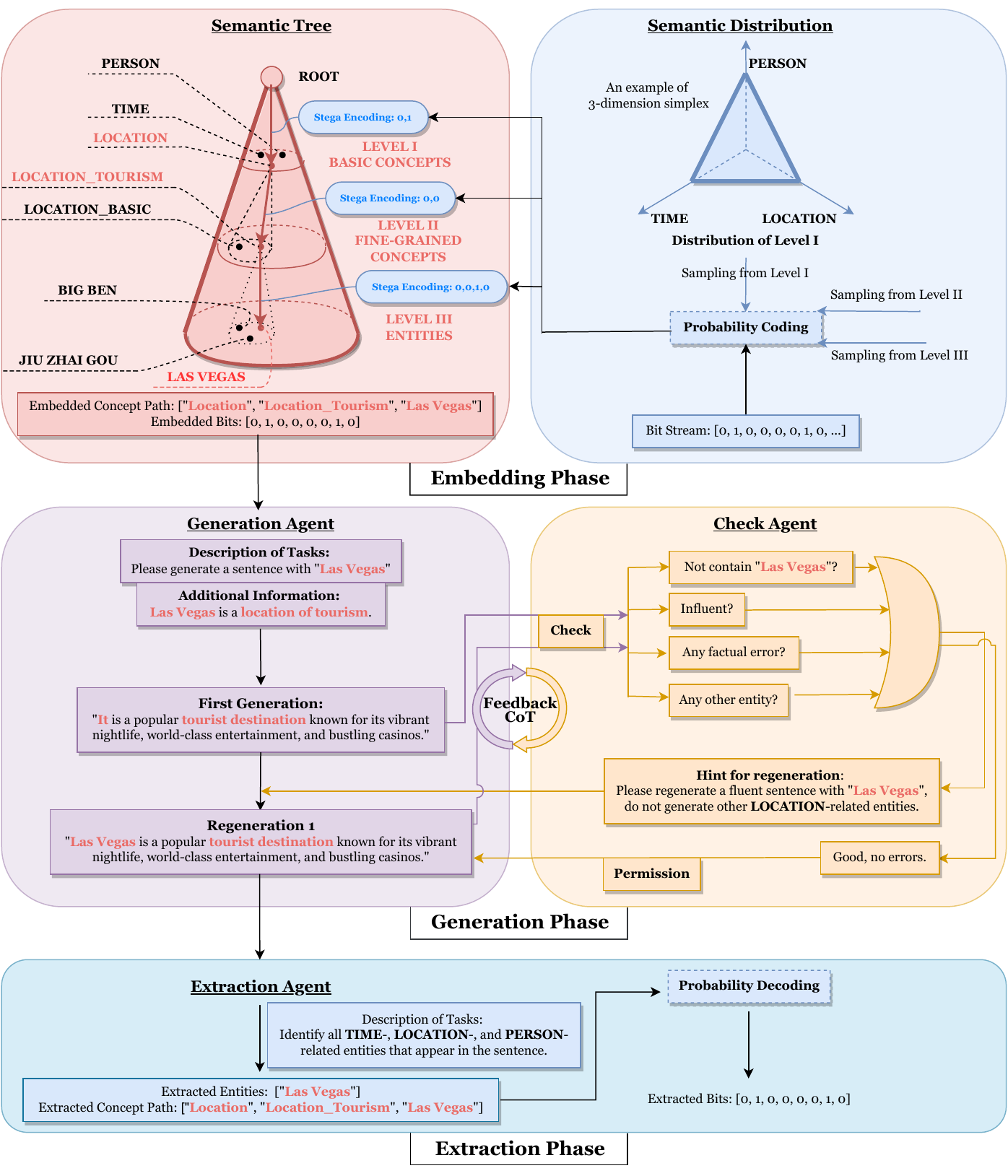}
\caption{Workflow of our framework, with a simple example. A {\bf type} ``{\bf Las Vegas$^1$}'' is selected according to the secret bit stream. The LLMs follow the instructions and generate a sentence ``\textbf{Las Vegas} is a popular \textbf{tourist destination} known for its vibrant
nightlife, world-class entertainment, and bustling casinos. '', which belongs to the {\bf class} of  {\bf type}  ``{\bf Las Vegas$^1$}''. The number of embedded bits depends on the estimated probability of corresponding entity, which is not manually set.}
\label{fig:Workflow}
\vspace{-0.5cm}
\end{figure*}
This section discusses a secure method of sampling from the semantic space.

For provably secure symbolic steganography methods such as METEOR \cite{METEOR}, MEC \cite{MEC}, and DISCOP \cite{DISCOP}, it is expected that the model-generated stegos are indistinguishable from the model-generated covers.
That means $D_{KL}(p(cover)||p(stego)) = 0$ \cite{Cachin}. To ensure the KL divergence is 0, secure sampling methods are often designed. As our method does not alter the sampling strategy of LLM, the stegos remain the same as the covers.

Although there is no difference between stegos and covers from a symbolic perspective, there is still a sampling issue from a semantic perspective. To begin with, we need to consider the empirical semantic distribution.

As the semantic space is made up of classes that represent different types, we must first extract sentence types from a large corpus and then estimate their probability by their frequency. 
In cases where the entities within a sentence cannot be obtained, a prediction model can be used.
An empirical semantic distribution can then be constructed by either counting sentences or training a model. 

To sample from this distribution, randomized methods are necessary to ensure the security. Pseudo-random functions (PRFs) are commonly used to convert a secret bit stream into a pseudo-random bit stream that follows a uniform distribution. The definition of PRFs is as follows: $F_{key}: \{0,1\}^s \rightarrow \{0,1\}^s $ is PRF if for all probabilistic polynomial-time (P.P.T.) classifiers $C$ and $key$, 
\begin{equation}
    |\mathbb{P}(C^{F_{key}}(1^s) = 1) - \mathbb{P}(C^{\mathcal{O}}(1^s) = 1)| \leq \frac{1}{\text{poly}(s)}
\end{equation}
where $\mathcal{O}$ is an oracle that randomly generates bits and poly$(\cdot)$ denotes polynomials.

The first step in sampling is to use a key and a PRF to invert the ciphertext into a uniformly distributed bitstream. Then the problem is to map the bit stream into entities.
Since a uniformly distributed bit stream $B=[b_1,b_2,\cdot\cdot\cdot,b_{|B|}]$ can be mapped to a decimal $\hat{B} = \sum_{i = 1}^{|B|}2^{-i}b_i \sim Unif[0,1]$. Then Arithmetic Coding (AC) can be used to map the decimal $\hat{B}$ to an probability interval which represents a {\bf class} in semantic space. In the sampling procedure, we start with the ROOT of the ontology-entity tree and begin to sum up the probability of its children. If we find that the sum has exceeded the $\hat{B}$ after adding the probability of $k$-th child $p_k$, we will stop adding the rest children and choose the last added $k$-th child. This procedure dose not change the original distribution because $\hat{B}$ is uniformly random in $[0,1]$, and the probability of $\hat{B} \in [\sum_{i = 1}^k p_i, \sum_{i = 1}^{k+1} p_i]$ is $p_k$ itself. In the process of going down through a path in that tree, the probability sum will approximate $\hat{B}$. In the end, a leaf node will be sampled and we compute the binary form of probability sum. We should find that the binary form of probability sum and the original bit stream $B$ share the same binary prefix, and this shared prefix is the embedded bits in the whole procedure.
In our practice, the algorithms used for sampling are referenced in Alg. \ref{alg:assign} and \ref{alg:sample}. 
It is possible to sample entities one by one, and these entities are finally combined to form a {\bf type}, then let LLM generate a stego belonging to the {\bf class} that relates to this {\bf type} in semantic space. As for decoding, it is a simple reverse progress. In this way, secret bits can be sequentially embedded in nodes of the ontology-entity tree.

It is worthy to mention that the construction of tree will not change the embedding capacity of the steganography system. Because the upper nodes of the tree is handcrafted, and it is necessary to keep the probability of the entities (and their combinations) the same as the estimation in our dataset, in order to produce semantically near-indistinguishable texts. Therefore, the embedding capacity only correlates with the entropy of the estimated distribution of the entities (and their combinations). It is advisable to maintain the original distribution during the sampling procedure.

\subsection{Feedback CoT for Stego Generation}

A {\bf class} is chosen for LLMs to generate after sampling from the semantic distribution. However, making LLMs generate sentences that belong to the {\bf class} is not always successful. A rejection sampling method must be used for LLMs to generate correct sentences. 
Therefore, we proposed a method called {\bf feedback Chain of Thought(CoT)} to increase the success rate of generation.

In the generation process, the LLM for generating stegos is called Generation Agent (GA). To check whether the generated stego satisfies the sampled {\bf class}, another LLM called Check Agent (CA) is used. For each generation loop, CA will return a hint for regeneration or it will consider the generated sentence compliant and return the approval.
With the feedback from CA, GA is able to efficiently adjust the generated sentence and quickly converge to a correct version. 
Feedback CoT reduces the number of iterations and saves a lot of time in the experiment. A result about feedback CoT in section \ref{Quality of Stegos} shows that it is able to decrease the perplexity of generated stegos and reduce the times of regeneration.
\vspace{-0.3cm}
\subsection{Workflow of Our Framework}\label{sec:workflow}
As Fig.\ref{fig:Workflow} shown, our framework works in 3 phases.

(1) {\bf Embedding Phase}: With a secret bit stream and a provably secure probability coding method, we use Alg.\ref{alg:sample} to sample a {\bf type} from the semantic distribution. During the sampling process, the paths of entities that selected are preserved for the next phases.

(2) {\bf Generation Phase}: We use the paths of entities and a description of task for GA to generate a primitive stego. Then the feedback loop starts running. CA generates a hint for regeneration and GA is instructed by CA to correct the stego. Finally, CA confirms that stego meets the requirements and gives permission to proceed to the next step. 

(3) {\bf Extraction Phase}: An LLM named extraction agent(EA) is instructed to extract the {\bf type} of sentence. Since the {\bf type} represents an interval of probability $[l,h]$, the decoding involves computing bit stream $B \in \{0,1\}^n$ that satisfies
$\sum_{i = 1}^{n} b_i*2^{-i} \in [l,h]$ and $\sum_{i = 1}^{n} b_i*2^{-i} \pm b_n*2^{-n}\notin [l,h] $.

\section{Experiment \& Result}

We use ChatGLM2-6B and ChatGLM2-6B-int4 as agents. ChatGLM2-6B-int4 is a weaker version of ChatGLM2-6B, but this model is extremely fast and only uses 6GB of GPU RAM. 
AC \cite{AC}, METEOR \cite{METEOR} and DISCOP \cite{DISCOP} are baselines of our experiments. They are state-of-the-art steganography methods.
For our experiments, we used a server equipped with 4 $\times$ RTX 3090 GPUs. The experiments consist of 3 parts. First, we measured the quality of our stegos and compared them with stegos generated by baselines and model-generated covers by random sampling. Then we tested the robustness of our method and AC against attacks that ignore/preserve the semantics of the original sentence.

\subsection{Quality of Stegos}\label{Quality of Stegos}
\begin{table}[t]
    \centering
    \setlength{\tabcolsep}{3pt}
    
    \begin{tabular}{l|l|ccc}
    
    \toprule[1.5pt]
    \midrule
         Methods & Model & PPL & Dist-3 & GPT-4 Sc. \\\midrule
         \multirow{2}{*}{AC} & 6B & 2065.73 & 0.8024 & 5.6381\\
          & 6B-int4 & 2206.96 & 0.8009& 5.3290\\ \midrule
         \multirow{2}{*}{METEOR} & 6B	&2087.19	&0.7962	&5.7718\\
         & 6B-int4	&2461.97	&0.8179	&5.1051\\ \midrule
        \multirow{2}{*}{DISCOP}  & 6B	&2049.18	&0.8008	&5.5750\\ 
        & 6B-int4	&2524.63	&0.8317	&5.0178\\\midrule
         \multirow{2}{*}{RS} & 6B & 2027.34 & 0.8050 & 5.6419\\
         & 6B-int4 &  2085.65 & 0.8001 & 5.1578\\\midrule
         \multirow{2}{*}{Ours} & 6B & {\bf 869.79} & {\bf 0.8753} & {\bf 7.3624}\\
          & 6B-int4 & 855.70 & 0.8742 & 7.1527\\
         \midrule
    \bottomrule[1.5pt]
    \end{tabular}
    \caption{Linguistic quality of the generated texts. 6B and 6B-int4 stand for ChatGLM2-6B and its 4-bit quantified version.}
    \label{tab:quality}
\end{table}
\begin{table}[t]
    \centering
    \setlength{\tabcolsep}{5pt}
    \begin{tabular}{l|l|ccc}
    
    \toprule[1.5pt]
    \midrule
         \multirow{2}{*}{Methods }&\multirow{2}{*}{Model }& \multicolumn{2}{c}{ER} & \multirow{2}{*}{MSR} \\
         & & bit/sent. & bit/tok. & \\
         \midrule
         \multirow{2}{*}{AC} & 6B & 2.5695 & 0.1788 & 0.459\\
          & 6B-int4 & 3.6863 & 0.2648 & 0.376\\ \midrule
         \multirow{2}{*}{METEOR} & 6B	& 2.1547	& 0.1747	& 0.458 \\
         &6B-int4 & 3.5797 & 0.2558 & 0.384 \\ \midrule
         \multirow{2}{*}{DISCOP}  & 6B & 2.4573	& 0.1755	& 0.461 \\
         &6B-int4 & 3.6764 & 0.2662 & 0.371\\ \midrule
         \multirow{2}{*}{RS} & 6B & - & - & 0.463\\
         & 6B-int4 & - & - & 0.457\\ \midrule
         \multirow{2}{*}{Ours} & 6B & {\bf 28.5088} & 0.3958 & {\bf 0.893}\\
         &6B-int4 & 27.8945 & {\bf 0.4130} & 0.884\\
         \midrule
    \bottomrule[1.5pt]
    \end{tabular}
    \caption{Embedding rate (ER) and mission success rate (MSR) of AC, RS and ours.}
    \label{tab:ERMSR}
    \vspace{-0.5cm}
\end{table}
\begin{table*}[ht]
    \centering
    \setlength{\tabcolsep}{5pt}
    \begin{tabular}{lc|cccc|cccc}
    
    \toprule[1.5pt]
    \midrule
         \multicolumn{2}{c|}{\multirow{2}{*}{Methods}}& \multicolumn{4}{c|}{Random}& \multirow{2}{*}{Paraphrase} & \multicolumn{3}{c}{SC}\\
         & & {Insert} & {Delete} & Replace & Swap & & SNR=5 & SNR=15 & SNR=60\\
         \midrule
         
         \multirow{4}{*}{AC-6B} & \multicolumn{1}{c|}{$|T| = 1$} & 0.034 & 0.012 & 0.016 & 0.010 & 0 & 0 & 0 & 0.0021\\
         & \multicolumn{1}{c|}{$|T| = 2$} & 0.049 & 0.025 & 0.044 & 0.037 & 0 & 0 & 0 & 0.0021\\
         & \multicolumn{1}{c|}{$|T| = 3$} & 0.073 & 0.057 & 0.065 & 0.028 & 0 & 0 & 0 & 0.0021\\
         & \multicolumn{1}{c|}{$|T| = 4$} & 0.070 & 0.056 & 0.061 & 0.021 & 0 & 0 & 0 & 0.0021\\
         \midrule
         \multirow{4}{*}{AC-6B-int4} & \multicolumn{1}{c|}{$|T| = 1$} & 0.031 & 0.027 & 0.026 & 0.017 & 0 & 0 & 0 & 0.0051\\
         & \multicolumn{1}{c|}{$|T| = 2$} & 0.052 & 0.032 & 0.059 & 0.047& 0 & 0 & 0 & 0.0037\\
         & \multicolumn{1}{c|}{$|T| = 3$} & 0.065 & 0.053 & 0.052 & 0.038& 0 & 0 & 0 & 0.0003\\
         & \multicolumn{1}{c|}{$|T| = 4$} & 0.072 & 0.033 & 0.057 & 0.052& 0 & 0 & 0 & 0.0000\\
         \midrule
         \multirow{4}{*}{Ours-6B} & \multicolumn{1}{c|}{$|T| = 1$} & {\bf 0.933} & {\bf 0.837} & {\bf 0.840} & {\bf 0.848}& {\bf 0.4203} & {\bf 0.8364} & {\bf 0.8370} & {\bf 0.8446}\\
         & \multicolumn{1}{c|}{$|T| = 2$} & 0.874 & 0.704 & 0.701 & 0.712& 0.2869 & 0.7187 & 0.7819 & 0.7898\\
         & \multicolumn{1}{c|}{$|T| = 3$} & 0.852 & 0.625 & 0.627 & 0.619& 0.2340 & 0.6249 & 0.7243 & 0.7339\\
         & \multicolumn{1}{c|}{$|T| = 4$} & 0.814 & 0.577 & 0.555 & 0.561& 0.2111 & 0.5279 & 0.6683 & 0.6780\\
         \midrule
         \multirow{4}{*}{Ours-6B-int4} & \multicolumn{1}{c|}{$|T| = 1$} & {\bf 0.931} & {\bf 0.821} & {\bf 0.823} & {\bf 0.817}& {\bf 0.4175} & {\bf 0.7778} & {\bf 0.7836} & {\bf 0.8132}\\
         & \multicolumn{1}{c|}{$|T| = 2$} & 0.869 & 0.700 & 0.681 & 0.702& 0.2819 & 0.7601 & 0.7676 & 0.7764 \\
         & \multicolumn{1}{c|}{$|T| = 3$} & 0.832 & 0.608 & 0.607 & 0.605& 0.2273 & 0.5943 & 0.7321 & 0.7334\\
         & \multicolumn{1}{c|}{$|T| = 4$} & 0.774 & 0.545 & 0.542 & 0.513& 0.1827 & 0.5330 & 0.6599 & 0.6862\\
         \midrule
    \bottomrule[1.5pt]
    \end{tabular}
    \caption{Decoding success rates of AC and ours, under attacks that ignore/preserve semantics. $|T|$ represents the length of {\bf type}.}
    \label{tab:DSR1}
    \vspace{-0.5cm}
\end{table*}

\begin{figure*}[ht]
\centering
\subfigure[Bertscore after paraphrasing]{
\includegraphics[width=0.45\textwidth]{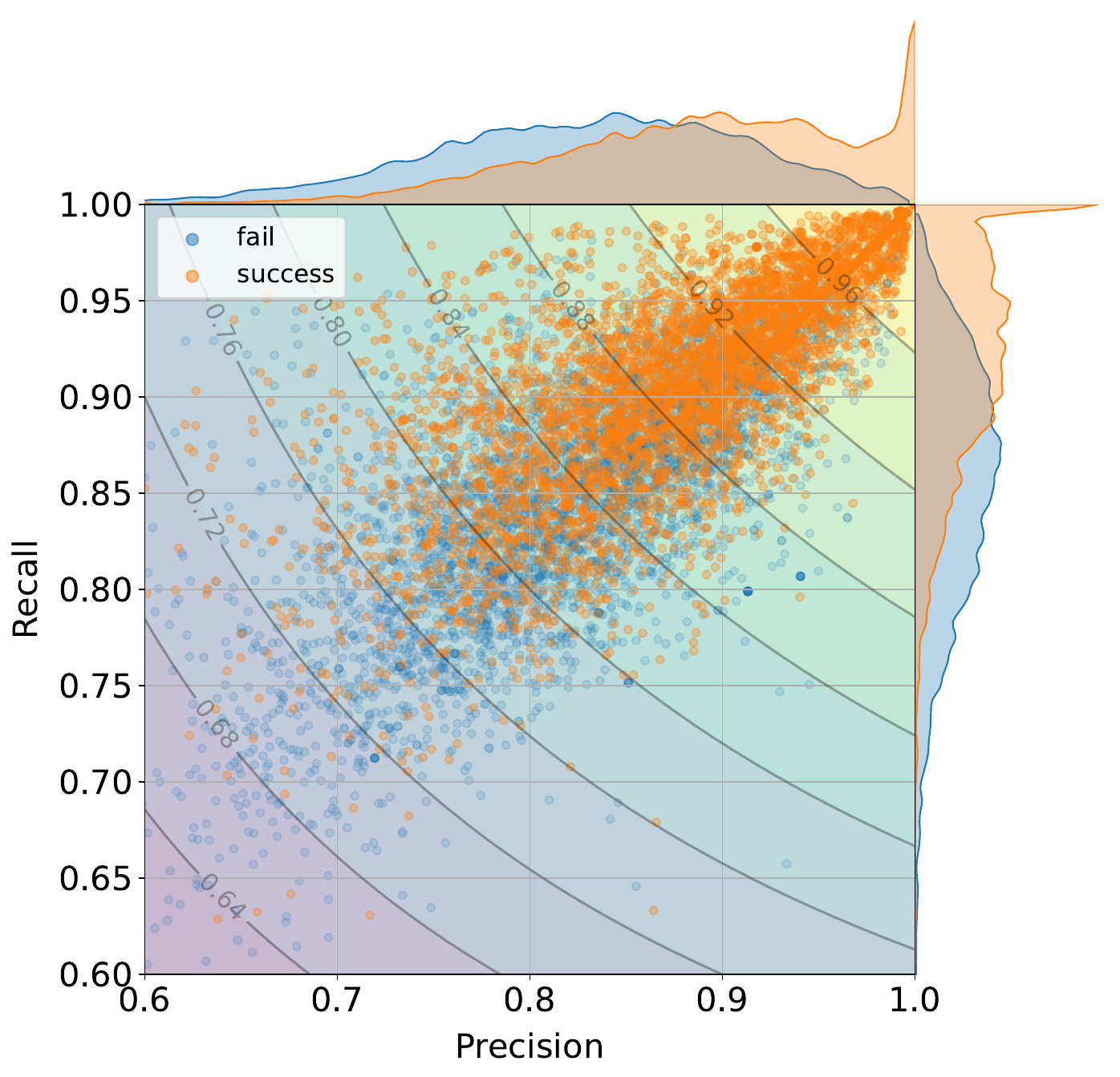}
\label{paraphrase}}
\subfigure[Bertscore after semantic communication (SNR=5)]{
\includegraphics[width=0.45\textwidth]{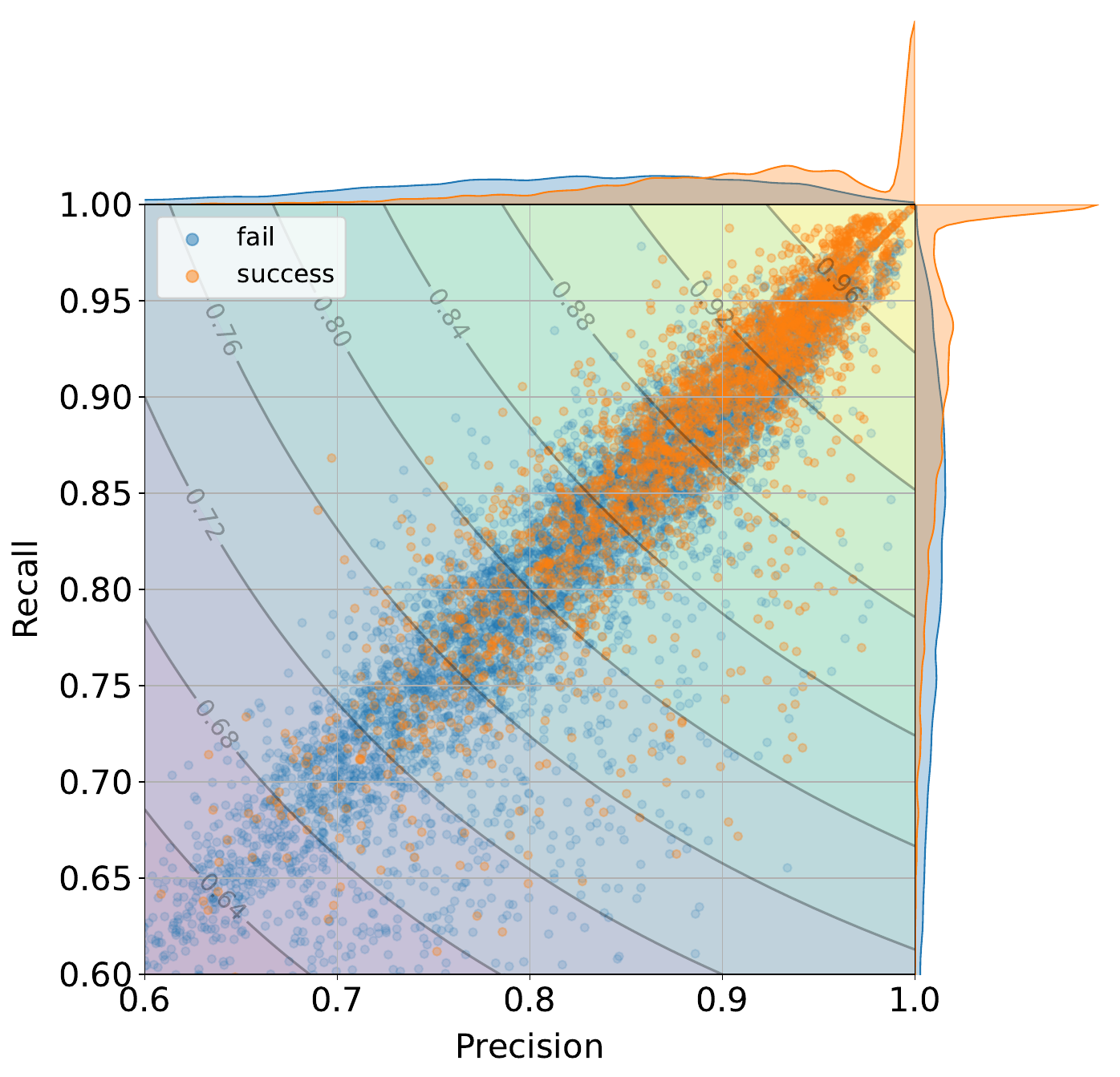}
\label{paraphrase}}
\caption{Bertscore \cite{bertscore} of stegos attacked by paraphrasing and semantic communication. The contour line in the middle represents F1 score, and the estimated marginal distributions of the two samples are plotted on the top and right sides.}
\label{fig:bertscore}
\end{figure*}

The linguistic quality of stegos is estimated by perplexity(PPL), distinct-$n$, and GPT-4 semantic rationality score. The PPL and distinct-$n$ are calculated as 
    $\text{PPL} = \exp\left(-\frac{1}{N}\sum_{i=1}^{N}\log p(x_i|\textbf{x}_{1:i-1})\right)$
and
$    \text{distinct-}n = \frac{count(unique\, ngrams)}{count(ngrams)}$.
PPL represents the fluency of stegos and distinct-$n$ measures the diversity.

The prompt used for GPT-4 to measure semantic rationality is:
\textit{You are a professional linguist, analyse the following sentences in terms of their semantic fluency and rationality and give them a score between 0 and 10.}

For this part of the experiment, we utilized ChatGLM2-6B/ChatGLM2-6B-int4 to generate text. AC \cite{AC} was employed to generate stegos, while the models were allowed to perform random sampling to generate covers. During generation, we set the top-$p$ truncation to 0.8 and the temperature to 0.8, following the generation configurations used by ChatGLM \cite{ChatGLM}. The results are presented in Tab.\ref{tab:quality}. 

Our framework generates stegos with a lower PPL than AC/RS. This is due to the CA checking the fluency of the stegos and providing prompts for the GA to regenerate. The feedback from CoT significantly improves the quality of the stegos. 

We also tested the embedding rate (ER) and the ``mission success rate'' (MSR), which indicates the probability of generated texts meeting the requirements in prompts. Details can be found in Tab.\ref{tab:ERMSR}.

Since the prompt is very restrictive on the model output, the entropy of the model predicted distribution is relatively low. This leads to the phenomenon that in some generated sentences, AC is not able to embed a single bit. However, such sentences are common in application scenarios. This result indicates that the entropy is compressed by LLMs and prompts with clear requests. The redundancy of the symbolic space has become difficult to use.

With the help of feedback CoT, the MSR of ours is about 2 times of AC and RS. The average number of loops in feedback CoT is 0.6312. 58.15\% of the sentences are allowed for output without regeneration and 29.25\%/8.31\%/4.22\% of the sentences require 1/2/3 iterations. Since the MSR of RS is 0.463, the MSR of the simplest rejection sampling with $n$ iterations can be estimated by
    $MSR_n = \sum_{i=1}^n 0.463 * (1-0.463)^{n-1} = 1-(1-0.463)^n$.
So to increase the MSR to 0.893, $n$ is about 3.5945. Feedback CoT can reduce the number of iterations to 17\% of the simplest rejection sampling.

\subsection{Robustness Against Attacks that Ignore or Preserve Semantics}

In this section, we first test the robustness of our method and AC against attacks that ignore semantics. These attacks include random insert/delete/replace/swap tokens in a sentence. Then we test the robustness of our method and AC against attacks that preserve semantics. These attacks include paraphrasing and semantic communication (SC) \cite{DEEPSC,MEM-DEEPSC}. These attacks completely change all of the tokens, but they have a probability to preserve the meaning of the original sentence. Details of these attacks are given below:
\textbf{Random Insert}: copy a random token from the sentence and insert it at a random position.
\textbf{Random Delete}: delete a random token.
\textbf{Random Replace}: replace a random token from the sentence with another random token.
\textbf{Random Swap}: pick two random tokens from the sentence and swap them.
\textbf{Paraphrase}: the prompt we used for GPT-4 to paraphrase is: \textit{You are an excellent editor. Rewrite the following sentences, keeping them about the same length and leaving the semantics as unchanged as possible.}
\textbf{Semantic communication}: semantic communication methods \cite{DEEPSC,MEM-DEEPSC} are designed to overcome the extremely high noise level. These methods have a probability of transmitting the correct meaning of sentences instead of the correct symbols.

Results are shown in Tab.\ref{tab:DSR1}.
The probability distribution predicted by LLM is changed so that AC's decoding is a disaster. In most cases, AC cannot decode the correct secret bits, and in most of these surviving examples, the attacks are targeted at the end of the sentence.  Therefore, the prefix of the decoded bits is likely to be the same as the encoded bits, which will be judged as success.
In contrast, our stegos shows explicit robustness against these attacks. Since secret bits are embedded in entities, attacks that randomly change tokens have a relatively low probability of destroying these entities. In some cases, the tokens that denote entities are changed, but the LLMs are able to correctly extract entities from perturbed tokens. This part of the robustness depends on the ability of the LLMs to correct sentences. 

Paraphrasing and SC completely change the sentences. As mentioned before, when the tokens are changed and the model prediction is different, AC is unable to decode. Our stegos retain some robustness against paraphrasing, and more than half can be decoded under SC.

However, paraphrasing and SC seem to subtly change the semantics. We measure the BertScore \cite{bertscore} of our stegos and paraphrasing/SC stegos to clarify the semantic noise level.

As shown in Fig. \ref{fig:bertscore}, the samples that could be decoded correctly are concentrated in the high-F1 region. In the paraphrased samples whose Bertscore F1 is more than 0.8/0.9, the decoding success rate is more than 60\%/75\%. The statistics of semantic communication in the same situation is 85\%/90\%. 
The result shows that most of our stegos remain robust under attacks that preserve semantics well. 

\section{Conclusion}

In this paper, we propose a semantic steganography framework based on LLMs. We use entities to build the semantic space with the help of ontology-entity tree, leverage Feedback CoT for rejection sampling, and apply AC for efficient encoding and decoding. Experiments showe that our framework are robust against attacks that ignore or preserve semantics. The embedding capacity of our framework is much higher than traditional symbolic steganography, while the quality of generated text is also better. 
Since our framework is able to work with black-box LLMs' API, it is easy to apply our method to construct covert communications in the real world scenario. 

\section{Limitations}


Our framework is difficult to operate in a low semantic-level entropy condition, which is different from the symbolic-level entropy. When the entities and relations in a sentence are fixed, like given the prompt ``1+1='', there is no redundancy to embed bits because the answer is just ``2''.

The LLM used for this framework will affect the quality and robustness of the stegos. Therefore, we recommend using those large LLMs with open APIs. However, if the local use of LLMs is a necessity, the need for GPU resources becomes a limitation. In our experiments, we used ChatGLM-6B-int4, which requires a maximum of 6GB of GPU RAM. Calculated as the product of GPU memory and time in use, generating a sentence takes about 12.0754 GB$\cdot$s. 

\section{Ethics Statement}

We propose a steganography framework based on LLMs. Due to the convenience of accessing LLMs, this method may have an impact on the security of LLMs generated texts. In our future work, we will study the detection method against LLMs generated steganographic texts. In our implementation and experiments we follow the licence of the used scientific artifacts.

\bibliography{anthology,custom}

\begin{thebibliography}{23}
\expandafter\ifx\csname natexlab\endcsname\relax\def\natexlab#1{#1}\fi

\bibitem[{Cachin(1998)}]{Cachin}
Christian Cachin. 1998.
\newblock An information-theoretic model for steganography.
\newblock In \emph{Information Hiding}, pages 306--318, Berlin, Heidelberg. Springer Berlin Heidelberg.

\bibitem[{Dai and Cai(2019)}]{HC3}
Falcon~Z. Dai and Zheng Cai. 2019.
\newblock \href {http://arxiv.org/abs/1907.06679} {Towards near-imperceptible steganographic text}.

\bibitem[{de~Witt et~al.(2023)de~Witt, Sokota, Kolter, Foerster, and Strohmeier}]{MEC}
Christian~Schroeder de~Witt, Samuel Sokota, J.~Zico Kolter, Jakob Foerster, and Martin Strohmeier. 2023.
\newblock \href {http://arxiv.org/abs/2210.14889} {Perfectly secure steganography using minimum entropy coupling}.

\bibitem[{Devlin et~al.(2019)Devlin, Chang, Lee, and Toutanova}]{BERT}
Jacob Devlin, Ming-Wei Chang, Kenton Lee, and Kristina Toutanova. 2019.
\newblock \href {http://arxiv.org/abs/1810.04805} {Bert: Pre-training of deep bidirectional transformers for language understanding}.

\bibitem[{Ding et~al.(2023)Ding, Chen, Wang, Zhao, Zhang, and Yu}]{DISCOP}
Jinyang Ding, Kejiang Chen, Yaofei Wang, Na~Zhao, Weiming Zhang, and Nenghai Yu. 2023.
\newblock \href {https://doi.org/10.1109/SP46215.2023.10179287} {Discop: Provably secure steganography in practice based on "distribution copies"}.
\newblock In \emph{2023 IEEE Symposium on Security and Privacy (SP)}, pages 2238--2255.

\bibitem[{Du et~al.(2022)Du, Qian, Liu, Ding, Qiu, Yang, and Tang}]{ChatGLM}
Zhengxiao Du, Yujie Qian, Xiao Liu, Ming Ding, Jiezhong Qiu, Zhilin Yang, and Jie Tang. 2022.
\newblock Glm: General language model pretraining with autoregressive blank infilling.
\newblock In \emph{Proceedings of the 60th Annual Meeting of the Association for Computational Linguistics (Volume 1: Long Papers)}, pages 320--335.

\bibitem[{Kaptchuk et~al.(2021)Kaptchuk, Jois, Green, and Rubin}]{METEOR}
Gabriel Kaptchuk, Tushar~M. Jois, Matthew Green, and Aviel~D. Rubin. 2021.
\newblock \href {https://doi.org/10.1145/3460120.3484550} {Meteor: Cryptographically secure steganography for realistic distributions}.
\newblock In \emph{Proceedings of the 2021 ACM SIGSAC Conference on Computer and Communications Security}, CCS '21, page 1529–1548, New York, NY, USA. Association for Computing Machinery.

\bibitem[{PaddleNLP(2021)}]{paddlenlp}
PaddleNLP. 2021.
\newblock Paddlenlp: An easy-to-use and high performance nlp library.
\newblock \url{https://github.com/PaddlePaddle/PaddleNLP}.

\bibitem[{Qin et~al.(2023)Qin, Xie, and Tao}]{MEM-DEEPSC}
Zhijin Qin, Huiqiang Xie, and Xiaoming Tao. 2023.
\newblock \href {https://doi.org/10.1109/ICC45041.2023.10279664} {Mem-deepsc: A semantic communication system with memory}.
\newblock In \emph{ICC 2023 - IEEE International Conference on Communications}, pages 3854--3859.

\bibitem[{Radford et~al.(2019)Radford, Wu, Child, Luan, Amodei, and Sutskever}]{GPT2}
Alec Radford, Jeffrey Wu, Rewon Child, David Luan, Dario Amodei, and Ilya Sutskever. 2019.
\newblock Language models are unsupervised multitask learners.

\bibitem[{Shen et~al.(2020)Shen, Ji, and Han}]{SAAC}
Jiaming Shen, Heng Ji, and Jiawei Han. 2020.
\newblock Near-imperceptible neural linguistic steganography via self-adjusting arithmetic coding.

\bibitem[{Touvron et~al.(2023)Touvron, Martin, Stone, Albert, Almahairi, Babaei, Bashlykov, Batra, Bhargava, Bhosale, Bikel, Blecher, Ferrer, Chen, Cucurull, Esiobu, Fernandes, Fu, Fu, Fuller, Gao, Goswami, Goyal, Hartshorn, Hosseini, Hou, Inan, Kardas, Kerkez, Khabsa, Kloumann, Korenev, Koura, Lachaux, Lavril, Lee, Liskovich, Lu, Mao, Martinet, Mihaylov, Mishra, Molybog, Nie, Poulton, Reizenstein, Rungta, Saladi, Schelten, Silva, Smith, Subramanian, Tan, Tang, Taylor, Williams, Kuan, Xu, Yan, Zarov, Zhang, Fan, Kambadur, Narang, Rodriguez, Stojnic, Edunov, and Scialom}]{LLAMA}
Hugo Touvron, Louis Martin, Kevin Stone, Peter Albert, Amjad Almahairi, Yasmine Babaei, Nikolay Bashlykov, Soumya Batra, Prajjwal Bhargava, Shruti Bhosale, Dan Bikel, Lukas Blecher, Cristian~Canton Ferrer, Moya Chen, Guillem Cucurull, David Esiobu, Jude Fernandes, Jeremy Fu, Wenyin Fu, Brian Fuller, Cynthia Gao, Vedanuj Goswami, Naman Goyal, Anthony Hartshorn, Saghar Hosseini, Rui Hou, Hakan Inan, Marcin Kardas, Viktor Kerkez, Madian Khabsa, Isabel Kloumann, Artem Korenev, Punit~Singh Koura, Marie-Anne Lachaux, Thibaut Lavril, Jenya Lee, Diana Liskovich, Yinghai Lu, Yuning Mao, Xavier Martinet, Todor Mihaylov, Pushkar Mishra, Igor Molybog, Yixin Nie, Andrew Poulton, Jeremy Reizenstein, Rashi Rungta, Kalyan Saladi, Alan Schelten, Ruan Silva, Eric~Michael Smith, Ranjan Subramanian, Xiaoqing~Ellen Tan, Binh Tang, Ross Taylor, Adina Williams, Jian~Xiang Kuan, Puxin Xu, Zheng Yan, Iliyan Zarov, Yuchen Zhang, Angela Fan, Melanie Kambadur, Sharan Narang, Aurelien Rodriguez, Robert Stojnic, Sergey Edunov, and Thomas
  Scialom. 2023.
\newblock \href {http://arxiv.org/abs/2307.09288} {Llama 2: Open foundation and fine-tuned chat models}.

\bibitem[{Wang et~al.(2023)Wang, Wang, Chen, Ding, Zhang, and Yu}]{ICstega}
Xilong Wang, Yaofei Wang, Kejiang Chen, Jinyang Ding, Weiming Zhang, and Nenghai Yu. 2023.
\newblock \href {https://doi.org/10.1109/ICASSP49357.2023.10095722} {Icstega: Image captioning-based semantically controllable linguistic steganography}.
\newblock In \emph{ICASSP 2023 - 2023 IEEE International Conference on Acoustics, Speech and Signal Processing (ICASSP)}, pages 1--5.

\bibitem[{Wang et~al.(2020)Wang, Ke, Zheng, Huang, Jiang, Zhu, and Huang}]{LCCC}
Yida Wang, Pei Ke, Yinhe Zheng, Kaili Huang, Yong Jiang, Xiaoyan Zhu, and Minlie Huang. 2020.
\newblock A large-scale chinese short-text conversation dataset.
\newblock In \emph{Natural Language Processing and Chinese Computing}, pages 91--103, Cham. Springer International Publishing.

\bibitem[{Xie et~al.(2021)Xie, Qin, Li, and Juang}]{DEEPSC}
Huiqiang Xie, Zhijin Qin, Geoffrey~Ye Li, and Biing-Hwang Juang. 2021.
\newblock \href {https://doi.org/10.1109/TSP.2021.3071210} {Deep learning enabled semantic communication systems}.
\newblock \emph{IEEE Transactions on Signal Processing}, 69:2663--2675.

\bibitem[{Yang et~al.(2024)Yang, Wu, Yi, Feng, and Zhang}]{NewBinstega}
Tianyu Yang, Hanzhou Wu, Biao Yi, Guorui Feng, and Xinpeng Zhang. 2024.
\newblock \href {https://doi.org/10.1109/TDSC.2023.3247493} {Semantic-preserving linguistic steganography by pivot translation and semantic-aware bins coding}.
\newblock \emph{IEEE Transactions on Dependable and Secure Computing}, 21(1):139--152.

\bibitem[{Yang et~al.(2019{\natexlab{a}})Yang, Guo, Chen, Huang, and Zhang}]{HC2}
Zhong-Liang Yang, Xiao-Qing Guo, Zi-Ming Chen, Yong-Feng Huang, and Yu-Jin Zhang. 2019{\natexlab{a}}.
\newblock \href {https://doi.org/10.1109/TIFS.2018.2871746} {Rnn-stega: Linguistic steganography based on recurrent neural networks}.
\newblock \emph{IEEE Transactions on Information Forensics and Security}, 14(5):1280--1295.

\bibitem[{Yang et~al.(2019{\natexlab{b}})Yang, Guo, Chen, Huang, and Zhang}]{RNNstega}
Zhong-Liang Yang, Xiao-Qing Guo, Zi-Ming Chen, Yong-Feng Huang, and Yu-Jin Zhang. 2019{\natexlab{b}}.
\newblock \href {https://doi.org/10.1109/TIFS.2018.2871746} {Rnn-stega: Linguistic steganography based on recurrent neural networks}.
\newblock \emph{IEEE Transactions on Information Forensics and Security}, 14(5):1280--1295.

\bibitem[{Yang et~al.(2021{\natexlab{a}})Yang, Zhang, Hu, Hu, and Huang}]{HC1}
Zhong-Liang Yang, Si-Yu Zhang, Yu-Ting Hu, Zhi-Wen Hu, and Yong-Feng Huang. 2021{\natexlab{a}}.
\newblock \href {https://doi.org/10.1109/TIFS.2020.3023279} {Vae-stega: Linguistic steganography based on variational auto-encoder}.
\newblock \emph{IEEE Transactions on Information Forensics and Security}, 16:880--895.

\bibitem[{Yang et~al.(2021{\natexlab{b}})Yang, Zhang, Hu, Hu, and Huang}]{VAEstega}
Zhong-Liang Yang, Si-Yu Zhang, Yu-Ting Hu, Zhi-Wen Hu, and Yong-Feng Huang. 2021{\natexlab{b}}.
\newblock \href {https://doi.org/10.1109/TIFS.2020.3023279} {Vae-stega: Linguistic steganography based on variational auto-encoder}.
\newblock \emph{IEEE Transactions on Information Forensics and Security}, 16:880--895.

\bibitem[{Zhang et~al.(2021)Zhang, Yang, Yang, and Huang}]{semantic1}
Siyu Zhang, Zhongliang Yang, Jinshuai Yang, and Yongfeng Huang. 2021.
\newblock \href {https://doi.org/10.1109/LSP.2020.3042413} {Linguistic steganography: From symbolic space to semantic space}.
\newblock \emph{IEEE Signal Processing Letters}, 28:11--15.

\bibitem[{Zhang et~al.(2020)Zhang, Kishore, Wu, Weinberger, and Artzi}]{bertscore}
Tianyi Zhang, Varsha Kishore, Felix Wu, Kilian~Q. Weinberger, and Yoav Artzi. 2020.
\newblock \href {http://arxiv.org/abs/1904.09675} {Bertscore: Evaluating text generation with bert}.

\bibitem[{Ziegler et~al.(2019)Ziegler, Deng, and Rush}]{AC}
Zachary~M. Ziegler, Yuntian Deng, and Alexander~M. Rush. 2019.
\newblock \href {http://arxiv.org/abs/1909.01496} {Neural linguistic steganography}.

\end{thebibliography}
\bibliographystyle{acl_natbib}

\appendix

\section{An Example of Generation \& Regeneration}

As Sec.\ref{sec:workflow} presents, the generated stegos will be checked by CA and regenerate if necessary. With the entities ``many years'' and ``Mr. Kee'', an simple example of regeneration is shown below:

\textit{\textbf{Prompt:}  Write a sentence containing the following keywords: many years, Mr Kee. In this case, many years is the element of time category, and no other time-like element should appear in the sentence. Mr Kee is an element in person category, no other person-like element should appear in the sentence.}

\textit{\textbf{Generation v0:} Mr. Kee dedicated many years to teaching the children of the village.}

\textit{\textbf{Feedback of CA:} ``Children'' should not be in the sentence because it is an element in person category.}

\textit{\textbf{Regeneration v1:} Mr. Kee toiled for many years on his beloved farm.}

\textit{\textbf{Feedback of CA:} ``Farm'' should not be in the sentence because it is an element in location category.}

\textit{\textbf{Regeneration v2:} Mr. Kee fostered wisdom over many years.}

\textit{\textbf{Feedback of CA:} Good. No errors.}

It is important to mention that the completeness of entities affects the rationality of stegos. Since we decide that some entities should be in the sentence and the others should not, the sentence might be somewhat strange if the chosen entities are not related.

\section{An Example of Paraphrase \& Semantic communication}

We trained the model proposed by \cite{DEEPSC} in Chinese corpus. The impact of paraphrasing and semantic communication is presented below.

\textit{\textbf{Entities:} \begin{CJK}{UTF8}{gbsn} 演员,观众\end{CJK}}

\textit{\textbf{Stego:}  \begin{CJK}{UTF8}{gbsn}
电影中的群众{\bf 演员}为影片增色不少，他们的不懈努力得到了观众的高度认可。
\end{CJK}}

\textit{\textbf{Paraphrased stego:}  \begin{CJK}{UTF8}{gbsn}
影片里的临时演员为电影增添了丰富的色彩，他们孜孜不倦的付出赢得了广大观众的赞誉。
\end{CJK}}

\textit{\textbf{Stego after semantic communication(SNR=5):}  \begin{CJK}{UTF8}{gbsn}
电影中的阮演员为诲真相产生了不少，他们的弹努力得到了观众的高度认可认可。
\end{CJK}}

\textit{\textbf{Stego after semantic communication(SNR=15):}  \begin{CJK}{UTF8}{gbsn}
中的群众演员为多个灵感色不少，他们的不懈努力得到了观众的高度认可。
\end{CJK}}

\textit{\textbf{Stego after semantic communication(SNR=60):}  \begin{CJK}{UTF8}{gbsn}
中的群众演员为多个冲突色不少，他们的不懈努力得到了观众的高度认可。
\end{CJK}}

In this case, the entities \begin{CJK}{UTF8}{gbsn}演员 and 观众\end{CJK} are not changed after paraphrase. Even those decoding methods based on retrieving are able to decode the correct bits. 

Although the sentence attacked by semantic communication appears to make no sense, it still contains the correct entities and can be decoded appropriately.

\newpage
\section{Additional Experiments}
\begin{table}[ht]
    \centering
    \begin{tabular}{c|ccccc}
    \toprule
    \hline
Method	&PPL	&Dist-3	&bit/sent	&bit/tok	&MSR \\\hline
Ours &869.79 &0.8753 &28.5088 &0.3958 &0.893\\
METEOR	&2461.97	&0.8179	&2.1547	&0.1747	&0.458 \\
DISCOP	&2524.63	&0.8317	&2.4573	&0.1755	&0.461 \\
\hline
\bottomrule
    \end{tabular}
    \caption{Comparison with METEOR\cite{METEOR} and DISCOP\cite{DISCOP}, testing by ChatGLM2-6B.}
    \label{tab:4}
\end{table}

\begin{table}[ht]
    \centering
    \begin{tabular}{c|ccccc}
    \toprule
    \hline
Method	&PPL	&Dist-3	&bit/sent	&bit/tok	&MSR \\\hline
AC	&2065.73	&0.8024	&2.5695	&0.1788	&0.459 \\
AC+prompt1	&3297.14	&0.9315	&3.9218	&0.2673	&0.326\\
AC+prompt2	&3622.07	&0.9567	&4.5410	&0.3154	&0.375\\
\hline
\bottomrule
    \end{tabular}
    \caption{Comparison with prompts that will cause high-entropy responses, testing by ChatGLM2-6B. Prompt 1 is \textit{You are a creative writter.} Prompt 2 is \textit{Please give a high entropy response.}}
    \label{tab:my_label}
\end{table}

Here we provide some additional results. We compared our method with 2 provably secure method METEOR \cite{METEOR} and DISCOP \cite{DISCOP}. Table \ref{tab:4} shows that METEOR and DISCOP also produce high-PPL texts and low MSR. These phenomenon is similar to the results of AC. In conclusion, METEOR, DISCOP and AC are symbol-level embedding methods and they are not suitable for LLMs due to the redundancy of symbol space is compressed.

If we want the LLMs to generate texts with high entropy, some prompts may be helpful. We tested 2 prompts: \textit{You are a creative writter.} and \textit{Please give a high entropy response.} These prompts significantly increase the embedding capacity of AC, and also degrade the quality of generated texts. However, the increased capacity is still much less than our method, and the PPL is unacceptably high. Moreover, the MSR of texts is also decreased. Therefore, using prompts to increase the capacity may be restricted by the quality  and MSR of texts.

\end{document}